\documentclass[twocolumn,aps,prc,showpacs,superscriptaddress,preprintnumbers,floatfix,nofootinbib]{revtex4}
\usepackage{epsfig,graphics}
\usepackage{graphicx}% Include figure files
\usepackage{dcolumn}% Align table columns on decimal point
\usepackage{bm}% bold math
\usepackage{amsmath}
\usepackage[usenames]{color}
\usepackage{ulem} %% for strike-through
\usepackage[colorlinks,linkcolor=blue,urlcolor=blue,citecolor=blue]{hyperref}

\usepackage{CJK}

\begin{document}

\begin{CJK*} {UTF8} {gbsn}

\title{Anisotropy fluctuation and correlation in central $\alpha$-clustered $^{12}$C+$^{197}$Au collisions}

\author{L. Ma}
\email[]{malong@fudan.edu.cn}
\affiliation{Key Laboratory of Nuclear Physics and Ion-beam Application (MOE), Institute of Modern Physics, Fudan University, Shanghai 200433, China}
\author{Y. G. Ma}
\email[]{mayugang@fudan.edu.cn}
\affiliation{Key Laboratory of Nuclear Physics and Ion-beam Application (MOE), Institute of Modern Physics, Fudan University, Shanghai 200433, China}
\affiliation{Shanghai Institute of Applied Physics, Chinese Academy of Sciences, Shanghai 201800, China}
\author{S. Zhang}
\email[]{song\_zhang@fudan.edu.cn}
\affiliation{Key Laboratory of Nuclear Physics and Ion-beam Application (MOE), Institute of Modern Physics, Fudan University, Shanghai 200433, China}

\date{}

\begin{abstract}

In the framework of a multi-phase transport model, fluctuation and correlation of the azimuthal anisotropies in $^{12}$C+$^{197}$Au collisions at $\sqrt{S_{NN}}$ = 200 GeV are explored. Properties of the initial eccentricity and final harmonic flow fluctuation resulted from $^{12}$C+$^{197}$Au collisions with Woods-Saxon configuration and triangular $\alpha$-clustering configurations of $^{12}$C are investigated via scaled variance, skewness and kurtosis. Comparisons are made between results from $\alpha$-clustered configurations and Woods-Saxon configuration. The triangular flow fluctuation is found to have particular sensitivity in distinguishing triangular $\alpha$-clustering structure of $^{12}$C. Furthermore, correlations between initial eccentricities and final flow harmonics are studied with strong multiplicity dependence observed for the correlation functions.

\end{abstract}


\maketitle

\section{Introduction}

Anisotropic flow as a typical collective behavior of particles emitted was proved to be a good observable to study the equation-of-state of the hot and dense matter created in the early stage of ultra-relativistic heavy-ion collisions~\cite{Ollitrault1992,Voloshin2008,Kolb1999,Ackermann2001,Teaney2001,Romatschke2007,Luo2017}. The flow harmonics defined as $v_{n}$ ($n$ = 2,3..) which are the Fourier expansion coefficients of the azimuthal anisotropy of the final-state particles are suggested to be sensitive to not only the early partonic dynamics but also the transportation properties of the source evolution~\cite{STAR2004flow,STAR2013flow,PHENIX2011flow}. Theoretically, the hydrodynamic model and the multi-phase transport model have been widely used to make predictions and was successful in giving comparable descriptions on experimental measurements~\cite{Ulrich2005,Gale2013,Qiu20111441,Song2011,Song2017,Chen2004,HuangSL2020}.
 
The essential role of the initial collision geometry was realized when looking into the flow harmonics of different collision systems scaled by initial eccentricities~\cite{Alver200798,Voloshin2006}. Important information about the fluctuating initial conditions is believed to be transferred to the final flow anisotropy during source expansion. In the past two decades, significant efforts have been made on the studies of initial geometry fluctuation effect on the final flow observables~\cite{Voloshin2007695,Lacey2010,Lacey2014112,Alver201081,DerradideSouza2011,Ma2011106,Ma2014,Han2011,Wang2014}. In addition, experimental measurements of the event-by-event anisotropic flow fluctuation suggest the close correlation between flow fluctuation and the fluctuations of the initial geometry~\cite{Alver2010104,Agakishiev2011,Sorensen2007,Margutti2019}. It was realized that the flow fluctuation on the event-by-event basis provides good access to the initial eccentricity fluctuation, elucidating both source evolution and initial fluctuation properties~\cite{Andrade2006,Petersen2010,Ma2016}.

The nuclear cluster is one of the essential features of a nuclear system. Systematic studies on the $\alpha$ cluster in light nuclei have been performed for more than 40 years~\cite{Brink1970,VONOERTZEN2006,Freer2018,He2014,He2016,Kanada2019,Liu2018,HuangBS2017,HuangBS2020}. Recent studies suggest that the intrinsic structure of light nuclei can be captured by the ``snapshots'' made in relativistic nuclear collisions by colliding light nucleus on the heavy nuclei target~\cite{Wojciech2014}. As the final-state anisotropy inherits geometric information from the initial state, it was proposed that the nuclear clustering configurations can be explored by harmonic flow in the final stage of the nuclear collisions. Extensive theoretical studies have been performed on harmonic flow as a probe for the $\alpha$-clustering structure of light nuclei in nuclear collisions involving $^{12}$C or $^{16}$O as the projectile and it is found that the intrinsic nuclear structure, predetermined by the arrangement of the $\alpha$ clusters, leads to quantitative difference of the flow measurements~\cite{SZhang2017,SZhang2018,Guo2019,Xu2018,GuoCC2017}. In addition, flow fluctuation characterized by the ratio of the cumulant flow is also proposed to be an approach for investigating clustering in light nuclei~\cite{Piotr2014,Rybczynski2018}. Studying the effects of initial $\alpha$-clustering on flow fluctuation offers valuable information about the sensitivity of experimental measurement of flow fluctuation in probing the signatures of $\alpha$-clustering configurations in light nuclei. 

In this paper, we present systematic simulation study of the fluctuation and correlation properties of initial eccentricity and final flow harmonics for $^{12}$C+$^{197}$Au collision at 200 GeV with a multi-phase transport model.  Influence of $\alpha$-clustering on flow fluctuation and initial-final anisotropy correlation is investigated in particular. This paper is organized as follows: In Sec. II, model and research methods are described. In Sec. III, results and discussion are presented. In Sec. IV, a brief summary is given.

\section{Model and method} 

A multi-phase transport model (AMPT) is employed for studying $^{12}$C+$^{197}$Au collisions. The model consists of four main components: the initial condition, partonic interactions, hadronization, and hadronic interactions~\cite{Zhang2000,Lin2005}. Starting from Monte Carlo Glauber initial conditions, phase space distributions of minijet partons and soft string excitations are generated from the Heavy Ion Jet Interaction Generator (HIJING) model~\cite{Wang1991}. In the string melting scenario, both excited strings and minijet partons are decomposed into partons followed by elastic partonic scatterings. Scatterings among partons are treated according to a parton cascade model - Zhang's parton cascade (ZPC) model which includes parton-parton elastic scattering with cross sections obtained from the theoretical calculations~\cite{Zhang1997}.  After partons stop interacting with each other, a simple quark coalescence model is used to combine partons into hadrons. Partonic matter is then turned into  hadronic matter and the subsequential hadronic interactions are simulated using a relativistic transport model (ART) including both elastic and inelastic scattering descriptions for baryon-baryon, baryon-meson and meson-meson interactions~\cite{Li1995}. With properly choosing partonic scattering cross section, the AMPT model was successful in describing many experimental observations in heavy-ion collisions at RHIC and LHC energies~\cite{Xu2016,Zhou2015,Jin2018,WangH2019}. 

\begin{figure}[htbp]
\centering
\resizebox{7.0cm}{!}{\includegraphics{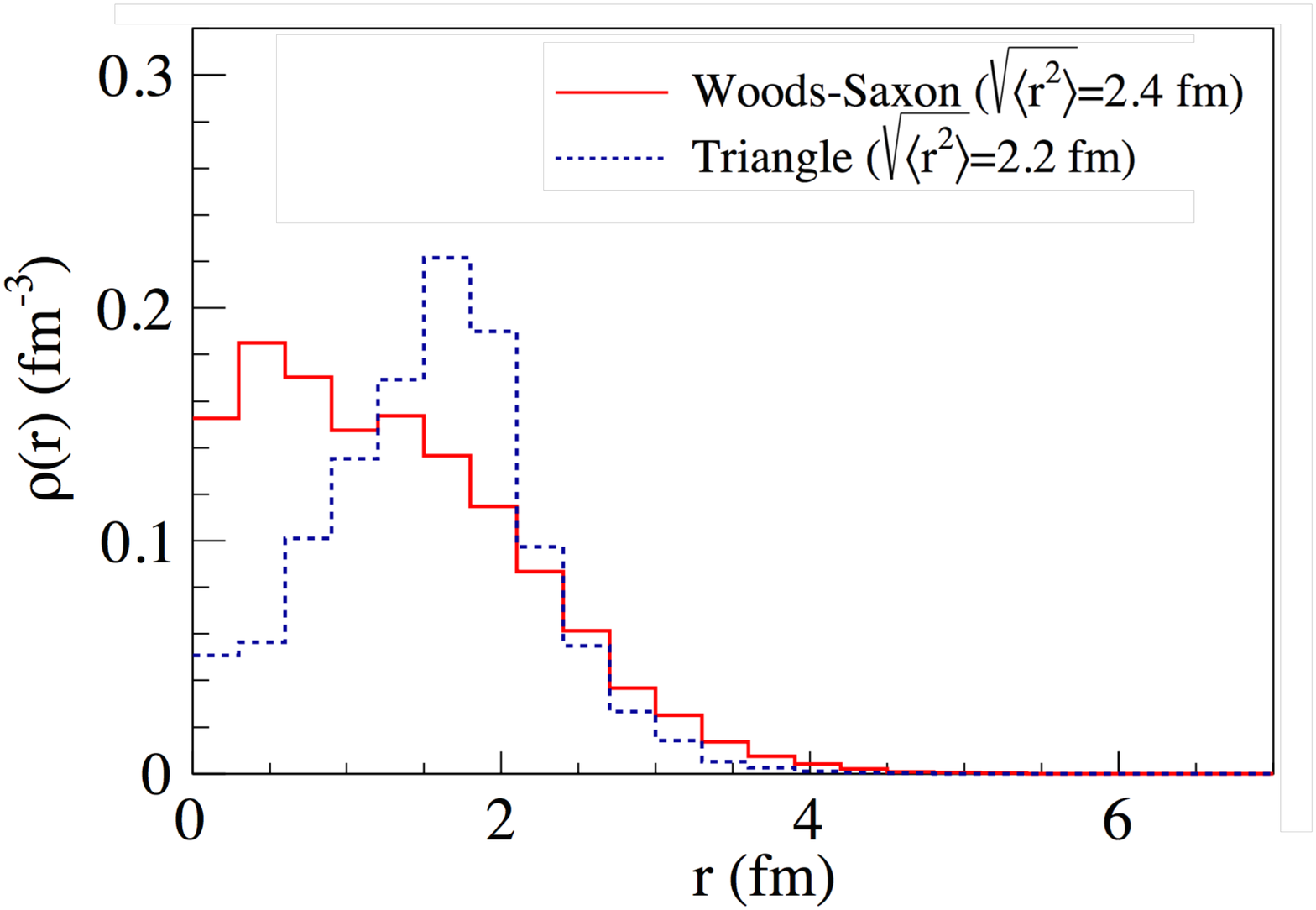}}
\caption{(Color online) Nucleon radial density distributions of $^{12}$C for the Woods-Saxon (W-S) and Triangle $\alpha$-clustering configurations. Results are normalized to the number of nucleons of $^{12}$C.}
\label{f1}
\end{figure}

Simulation events are generated for $^{12}$C+$^{197}$Au collisions at center-of-mass energy of 200 GeV. By default, the initial nucleon distribution in $^{12}$C and $^{197}$Au are initialized event-by-event according to Woods-Saxon distribution from the HIJING model. For studying the $\alpha$-clustering effect, $^{12}$C is configured with three $\alpha$ clusters in triangular structure based on the information given by an extended quantum molecular dynamics model (EQMD)~\cite{Maruyama1996}. The distribution of the radial centers of the $\alpha$ clusters are assumed to be Gaussian and initialized according to $e^{-0.5(\frac{r-r_{c}(\alpha)}{\sigma_{r_{c}}(\alpha)})^{2}}$, where $r_{c}(\alpha)$ is the averaged radial center coordinate of an $\alpha$ cluster and $\sigma_{r_{c}(\alpha)}$ is the width of the distribution. Nucleons inside each $\alpha$ cluster are initialized according to Woods-Saxon distribution. In our simulation study, the parameters for configuring the $^{12}$C are taken from the EQMD calculation with $r_{c}(\alpha)$ = 1.8 fm and $\sigma_{r_{c}}(\alpha)$ = 0.1 fm for the triangular $\alpha$-clustering configuration. Details about the methodology applied for the initialization of the collision system can be found in Refs.~\cite{He2014,He2016}. Fig.~\ref{f1} shows the comparison of the normalized nucleon radial density distributions initialized with Woods-Saxon and triangular $\alpha$-clustering configurations of $^{12}$C in this study. One can find that though the root mean square radius of the two configurations are comparable, the difference in the shape of the density distribution can be clearly seen.

\section{Results and Discussion}

\subsection{Eccentricity fluctuation in $^{12}$C+$^{197}$Au collisions}

For nuclear-nuclear collision, the initial geometric anisotropy of the collision zone in the transverse plane (perpendicular to the beam direction) can be characterized with eccentricity. It was argued that the magnitude and trend of the eccentricity and its fluctuation imply specifically testable predictions for final stage harmonic flow~\cite{Lacey2010,Drescher2007,Broniowski2007}. Definition of the eccentricity for the $n$th-order harmonic in the coordinate space of the participant nucleons or partons for a single collision event is in the form:

\begin{equation}
\varepsilon _{n}\left\{P\right\}  = \frac{{\sqrt {\left\langle {r^{n} \cos (n\varphi)} \right\rangle ^2 + \left\langle {r^{n} \sin (n\varphi)}
\right\rangle ^2 } }}{{\left\langle {r^{n} } \right\rangle }},
\label{q1}
\end{equation}
where $r$ and $\varphi$ are position and azimuthal angles of each nucleon or parton in the transverse plane. $\varepsilon _{n}\left\{P\right\}$ characterizes the eccentricity through the distribution of participant nucleons or partons which naturally contains event-by-event fluctuation. $\varepsilon _{n}\left\{P\right\}$ defined in this way is usually named as a ``participant eccentricity''.  

Initial eccentricity can also be quantified by cumulants of $\varepsilon _{n}\left\{P\right\}$ following the same way as in Ref.~\cite{Miller2003}. The definitions in terms of two-particle and four-particle cumulant moments of the eccentricities are in the form:

\begin{equation}
\begin{split}
&\varepsilon_{n}\left\{2\right\} = \sqrt{ \left\langle{\varepsilon^{2}_{n}\left\{P\right\}}\right\rangle },\\
&\varepsilon_{n}\left\{4\right\} = ( 2\left\langle{\varepsilon^{2}_{n}\left\{P\right\}}\right\rangle^{2}-\left\langle{\varepsilon^{4}_{n}\left\{P\right\}}\right\rangle )^{1/4}.
\end{split}
\label{q2}
\end{equation}

\begin{figure*}[htbp]
\centering
\resizebox{16.0cm}{!}{\includegraphics{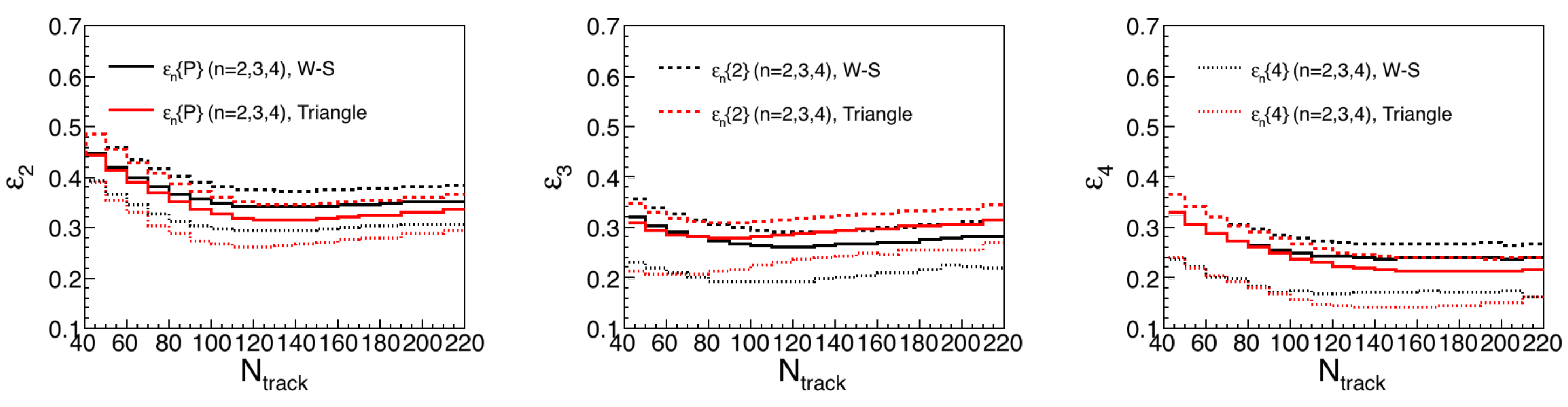}}
\caption{(Color online) Initial eccentricity $\varepsilon_{n}$ (n=2,3,4) as a function of $N_{track}$ for $^{12}$C+$^{197}$Au collisions at $\sqrt{S_{NN}}$ = 200 GeV in AMPT model. Results from the Triangle $\alpha$-clustering configuration are compared with Woods-Saxon (W-S) configuration without $\alpha$ clusterization.}
\label{f2}
\end{figure*}

\begin{figure*}[htbp]
\centering
\resizebox{16.0cm}{!}{\includegraphics{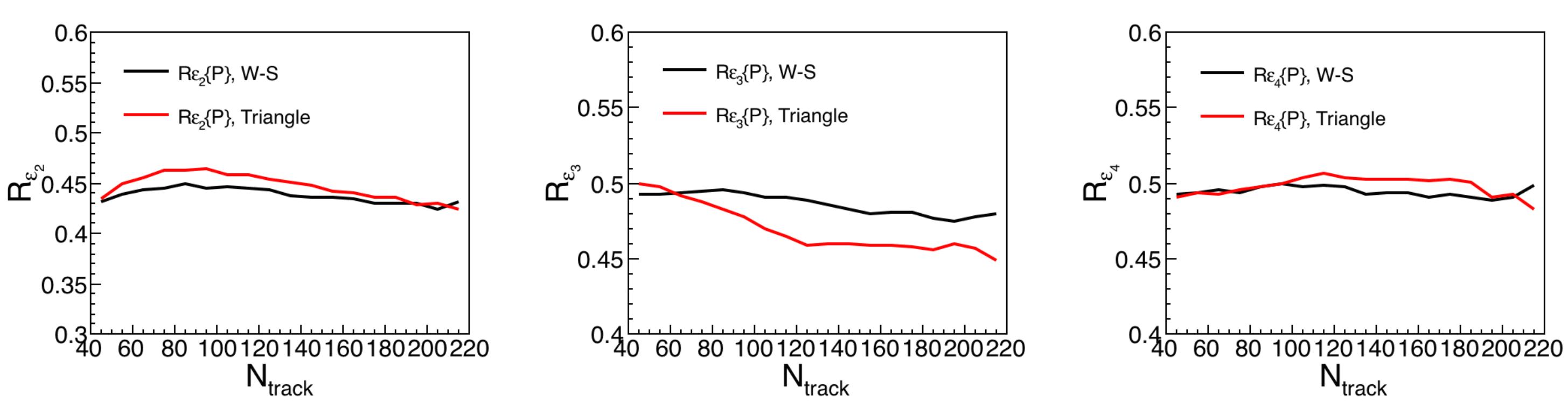}}
\caption{(Color online) Eccentricity fluctuation $R_{\varepsilon_{n}}$ as a function of $N_{track}$ for $^{12}$C+$^{197}$Au collisions at $\sqrt{S_{NN}}$ = 200 GeV in AMPT model. Results from the Triangle clustering configuration are compared with Woods-Saxon (W-S) configuration without $\alpha$ clusterization. Panels (a-c) show scaled standard deviations $R_{\varepsilon_{n}}$ for different order of harmonics.}
\label{f3}
\end{figure*}

In central nuclear collision involving clustered light nuclei colliding on a heavy-ion target, the profile of the initial collision zone inherits intrinsic geometry of the nucleon distribution of the light nuclei. The clustering configuration of nuclei is expected to be reflected by the eccentricity coefficients. In Fig.~\ref{f2}, simulation results of $\varepsilon_{n}$ are shown up to fourth order of the harmonic as a function of $N_{track}$ for central $^{12}$C+$^{197}$Au collisions with different $\alpha$-clustering configurations of $^{12}$C. $N_{track}$ is defined to be the number of charged particles within rapidity window, 1.0 $<$ y $<$ 1.0 and transverse momentum window 0.2 $<$ $p_{T}$ $<$ 6.0 GeV/c in the collision event. $\varepsilon_{n}$ results from two- and four-particle cumulant definitions by Eq.~\ref{q2} are compared with participant definition by Eq.~\ref{q1}. It is found $\varepsilon_{n}\left\{P\right\}$ (n=2,3,4) for all the configurations are quantitatively between $\varepsilon_{n}\left\{2\right\}$ and $\varepsilon_{n}\left\{4\right\}$ over the whole $N_{track}$ range. Non-monotonous dependence on $N_{track}$ can be seen for $\varepsilon_{n}$ resulted from the Triangle and unclustered Woods-Saxon (W-S) configurations. In comparison, Triangle and W-S configurations exhibit clear decreasing trend for $N_{track}$ $<$ 100. $\varepsilon_{3}$ and $\varepsilon_{4}$ are similar in trend as a function of $N_{track}$ but show different magnitude orderings for different $\alpha$-clustered cases. 
 
Event-by-event fluctuation of the eccentricity is studied by first looking into the scaled standard deviations (scaled variance) of eccentricity distribution defined as:

\begin{equation}
R_{\varepsilon_{n}}=\sigma_{\varepsilon_{n}}/\langle \varepsilon_{n} \rangle =\sqrt{\frac{\langle \varepsilon^{2}_{n} \rangle - \langle \varepsilon_{n} \rangle^{2}}{\langle \varepsilon_{n} \rangle^{2}}},
\label{q3}
\end{equation}
where $\sigma_{\varepsilon_{n}}$ is the variance which quantifies the absolute fluctuation. The brackets denote event averaging. Fig.~\ref{f3} comparatively shows the scaled standard deviation of eccentricity fluctuation $R_{\varepsilon_{n}}$ (n=2,3,4) for W-S and $\alpha$-clustering configurations of $^{12}$C. In comparison, $R_{\varepsilon_{2}}$ and 
$R_{\varepsilon_{4}}$ show a similar decreasing trend as a function of $N_{track}$. This is consistent with the expectation that quadrangularity $\varepsilon_{4}$ is strongly correlated with ellipticity $\varepsilon_{2}$ and their fluctuations should behave similarly~\cite{Qiu201184}. It is interesting to see the difference between W-S and Triangle configurations that $R_{\varepsilon_{3}}$ shows different ordering compared with $R_{\varepsilon_{2}}$ and $R_{\varepsilon_{4}}$. It could be understood in view of the geometrical origin that triangular structure from the intrinsic geometric bias contributes more to the ellipticity or quadrangularity fluctuation but less triangularity fluctuation. It is also found that for both $\varepsilon_{n}$ and $R_{\varepsilon_{n}}$ the discrepancy between different configurations tend to converge when approaching low $N_{track}$ and become diverge at higher $N_{track}$.

In addition to the standard deviation, skewness and kurtosis which characterize the non-Gaussian fluctuation properties have been used to study flow fluctuation and explain possible contributions to the splitting of higher-order flow cumulants~\cite{Giacalone2017,Bhalerao2019}. The standardized skewness and kurtosis of $\varepsilon$ fluctuations are defined as:

\begin{equation}
\begin{split}
&S_{\varepsilon_{n}} = \frac{\langle(\varepsilon_{n}-\langle \varepsilon_{n} \rangle)^{3}\rangle}{\langle(\varepsilon_{n}-\langle \varepsilon_{n} \rangle)^{2}\rangle^{3/2}},\\
&K_{\varepsilon_{n}} = \frac{\langle(\varepsilon_{n}-\langle \varepsilon_{n} \rangle)^{4}\rangle}{\langle(\varepsilon_{n}-\langle \varepsilon_{n} \rangle)^{2}\rangle^{2}}-3
\end{split}
\label{q4}
\end{equation}
where angular brackets denote an average over events. Both skewness and kurtosis vanish if the distribution is Guassian.

\begin{figure*}[htbp]
\centering
\resizebox{16.0cm}{!}{\includegraphics{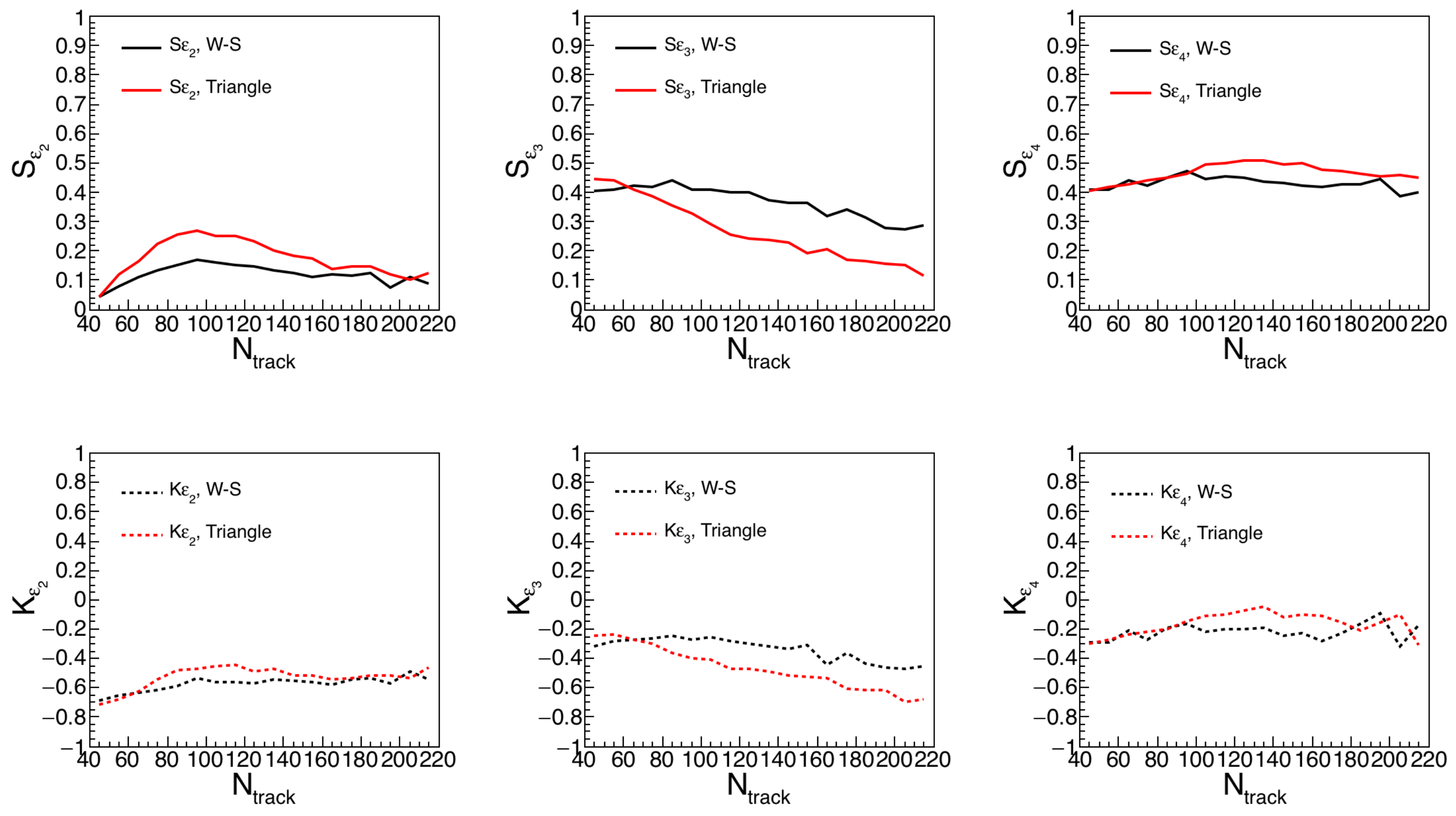}}
\caption{(Color online) The AMPT results on skewness and kurtosis of $\varepsilon_{n}$ (n=2,3,4) fluctuation for $^{12}$C+$^{197}$Au collisions at 200 GeV with $\alpha$-clustering and W-S configurations of $^{12}$C, as a function of $N_{track}$. Upper panels: Skewness of $\varepsilon_{n}$ fluctuation. Lower panels: Kurtosis of $\varepsilon_{n}$ fluctuation.}
\label{f4}
\end{figure*}

In addition to the scaled variance, we further studied skewness and kurtosis of the eccentricity fluctuation. Fig.~\ref{f4} displays skewness and kurtosis of $\varepsilon_{n}$ for W-S and triangle $\alpha$-clustering configurations of $^{12}$C. The skewness of eccentricity fluctuation $S_{\varepsilon_{n}}$ defined by Eq.(~\ref{q4}) takes a similar trend as $R_{\varepsilon_{n}}$ for a different order of harmonics. In comparison, $S_{\varepsilon_{2,4}}$ and $K_{\varepsilon_{2,4}}$ for triangle and W-S configurations show tiny multiplicity dependence. In addition, a significant difference in both magnitude and trend can be seen for $S_{\varepsilon_{3}}$ and $K_{\varepsilon_{3}}$ as a function of $N_{track}$ in the cases of different initial configurations. The pronounced dependencies of $N_{track}$ of skewness and kurtosis of $\varepsilon_{2}$ and $\varepsilon_{3}$ fluctuations imply possible approaches for studying $\alpha$ clusterization of the collision system by looking into the high order fluctuation properties.

\subsection{Flow fluctuation in $^{12}$C+$^{197}$Au collisions}

It is predicted that the azimuthal anisotropy characterized by anisotropic flow coefficients $v_{n}$ (n=2,3..) of final-state particle distribution is correlated in magnitude and phase with initial eccentricity. Considering the event-by-event fluctuation, calculation of $v_{n}$ can be done referred to the participant plane angle~\cite{Voloshin2007695} defined with the coordinate information of initial participant partons:

\begin{equation}
\psi_{n}\left\{PP\right\}  = \frac{1}{n}\left[ \arctan\frac{\left\langle {r^{2} \sin
(n\varphi_{PP})} \right\rangle}{\left\langle {r^{2} \cos (n\varphi_{PP})}
\right\rangle} + \pi \right],
 \label{q5}
\end{equation}
where $n$ denotes the $n$th-order participant plane,  $r$ and $\varphi_{PP}$ are the position and azimuthal angle of each parton in AMPT initial stage and the bracket denotes density weighted average. Flow coefficients $v_{n}$ calculated with respect to the participant plane $\psi_{n}$ are defined as

\begin{equation}
v_{n}\left\{PP\right\} = \left\langle cos[n(\phi-\psi_{n}\left\{PP\right\})] \right\rangle,
\label{q6}
\end{equation}
where $\phi$ is azimuthal angle of final-state charged hadrons in the momentum space, and the average $\langle \cdots\rangle$ denotes event average. Similar to the definition of the eccentricity, this method for calculation of $v_n$ is referred to as participant plane $v_n$ which was widely used for flow calculations in different models.  

One can also characterize the different orders of azimuthal anisotropies with Event plane (EP) method~\cite{Poskanzer1998,Voloshin2008}. Unlike participant plane which is not accessible experimentally, event plane can be reconstructed using final-state charged particles. Definition of the $n$th-order event plane angle is in the form:

\begin{equation}
\psi_{n}\left\{EP\right\}  = \frac{1}{n} \arctan\frac{\left\langle {\omega \sin(n\phi)} \right\rangle}{\left\langle {\omega \cos (n\phi)}\right\rangle},
 \label{q7}
\end{equation}
where $\phi$ and $\omega$ are azimuthal angle and weight for the final particle, respectively. Flow coefficients $v_{n}$ w.r.t the n-th order event plane $\psi_{n}\left\{EP\right\}$ is defined as:

\begin{equation}
v_{n}\left\{EP\right\} = \frac{\langle cos(n[\phi-\psi_{n}\left\{EP\right\}]) \rangle}{Res\left\{\psi_{n}\left\{EP\right\}\right\}},
\label{q8}
\end{equation}
where $Res\left\{\psi_{n}\left\{EP\right\}\right\}$ is the resolution of event plane angle and the brackets indicate average over particles.

\begin{figure*}[htbp]
\centering
\resizebox{15.0cm}{!}{\includegraphics{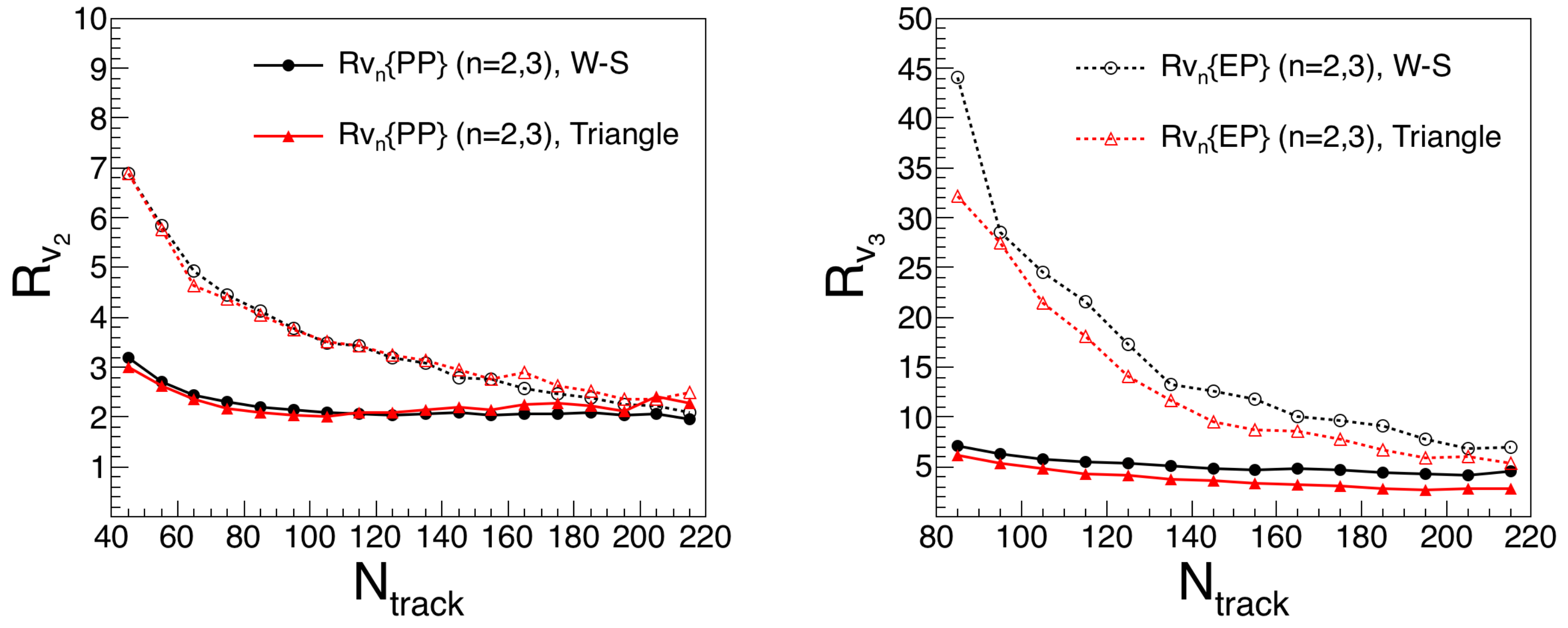}}
\caption{(Color online) The AMPT results on elliptic and triangular flow fluctuation as a function of $N_{track}$ for W-S and Triangle $\alpha$-clustering configurations of $^{12}$C in $^{12}$C+$^{197}$Au collisions at $\sqrt{s_{NN}} = 200 GeV$. Results are shown from both event plane method and participant plane methods. Left panel: $v_{2}$ fluctuation. Right panel: $v_{3}$ fluctuation.}
\label{f5}
\end{figure*}

Ideal hydrodynamics predicts linear response to the initial eccentricities of the final flow harmonics~\cite{Qiu201184}. Approximate proportionality of the flow coefficients $v_{n}$ to the eccentricities $\varepsilon_{n}$ is suggested to hold for n=2 and 3, $v_{n}$=$\kappa$$\varepsilon_{n}$ (n=2,3), where $\kappa$ is the linear response coefficient. Assuming linear flow response, flow fluctuation quantified with scaled standard deviation should be approximately equal to the eccentricity fluctuation quantified in the same way.

\begin{equation}
\frac{\sigma_{v_{n}}}{\langle v_{n} \rangle} \approx \frac{\sigma_{\varepsilon_{n}}}{\langle \varepsilon_{n} \rangle}
\label{q9}
\end{equation}

This relation can be applied only on the premise that initial-final correlation is dominated by linear response. Though fluctuations of $v_{n}$ mainly stem from the fluctuations of $\varepsilon_{n}$, during the source evolution, non-linear responses may play an important role in the development of final flow fluctuation with the presence of nonzero higher order effects~\cite{Yan2015,Renk2014,Retinskaya2014,Noronha2016,Bhalerao2011}. Nevertheless, we can still examine the flow fluctuation to see how sensitive it is in distinguishing initial nuclear clustering structure.

Defined in the same way as Eq.(~\ref{q3}), Fig.~\ref{f5} shows the simulation results of the scaled variance of the elliptic and triangular flow fluctuation of the charged hadrons at mid-rapidity (-1.0 $<$ y $<$ 1.0) for $^{12}$C+$^{197}$Au collisions at $\sqrt{s_{NN}}$ = 200 GeV. Comparisons are made between $\alpha$-clustered case and Woods-Saxon case where the nucleons are distributed without clusterization. $R_{v_{n}}\left\{EP\right\}$ which quantifies $v_{n}\left\{EP\right\}$ fluctuation are shown in addition to $v_{n}\left\{PP\right\}$ fluctuation $R_{v_{n}}\left\{PP\right\}$ based on the participant plane method. $R_{v_{n}}\left\{EP\right\}$ ($n$=2,3) show similar behavior as $R_{v_{n}}\left\{PP\right\}$ but comparably much higher for low-multiplicity collisions owing to the smearing effect brought by the source evolution. 

It is found that the scaled variances of both $v_{2}$ and $v_{3}$ present a monotonous decreasing trend at low $N_{track}$ ($N_{track}$$<$100) and change slowly towards higher $N_{track}$. $R_{v_{2}}$ is insensitive for distinguishing Traingle and W-S configurations as they are close in magnitude. A magnitude ordering can be seen for $R_{v_{3}}$ resulted from W-S and clustering configurations. It is observed that $R_{v_{3}}$ for Triangle configuration is lower than W-S configuration similar to the behavior of $R\varepsilon_{3}$ which is consistent with the picture of the flow response. 

Experimentally, one could take $R_{v_{n}}\left\{EP\right\}$ as a probe of $\alpha$-clustering structure in $^{12}$C. Similar to probing nuclear clustering structure with the ratio of flow harmonics, by choosing a isobar nucleus close to $^{12}$C with non-exotic structure colliding against $^{197}$Au as a reference, quantitative difference of $R_{v_{n}}\left\{EP\right\}$ between $\alpha$-clustering collision system and unclustered reference system can serve as a good probe to distinguish the clustering nuclear structure.

\begin{figure}[htbp]
\centering
\resizebox{7.0cm}{!}{\includegraphics{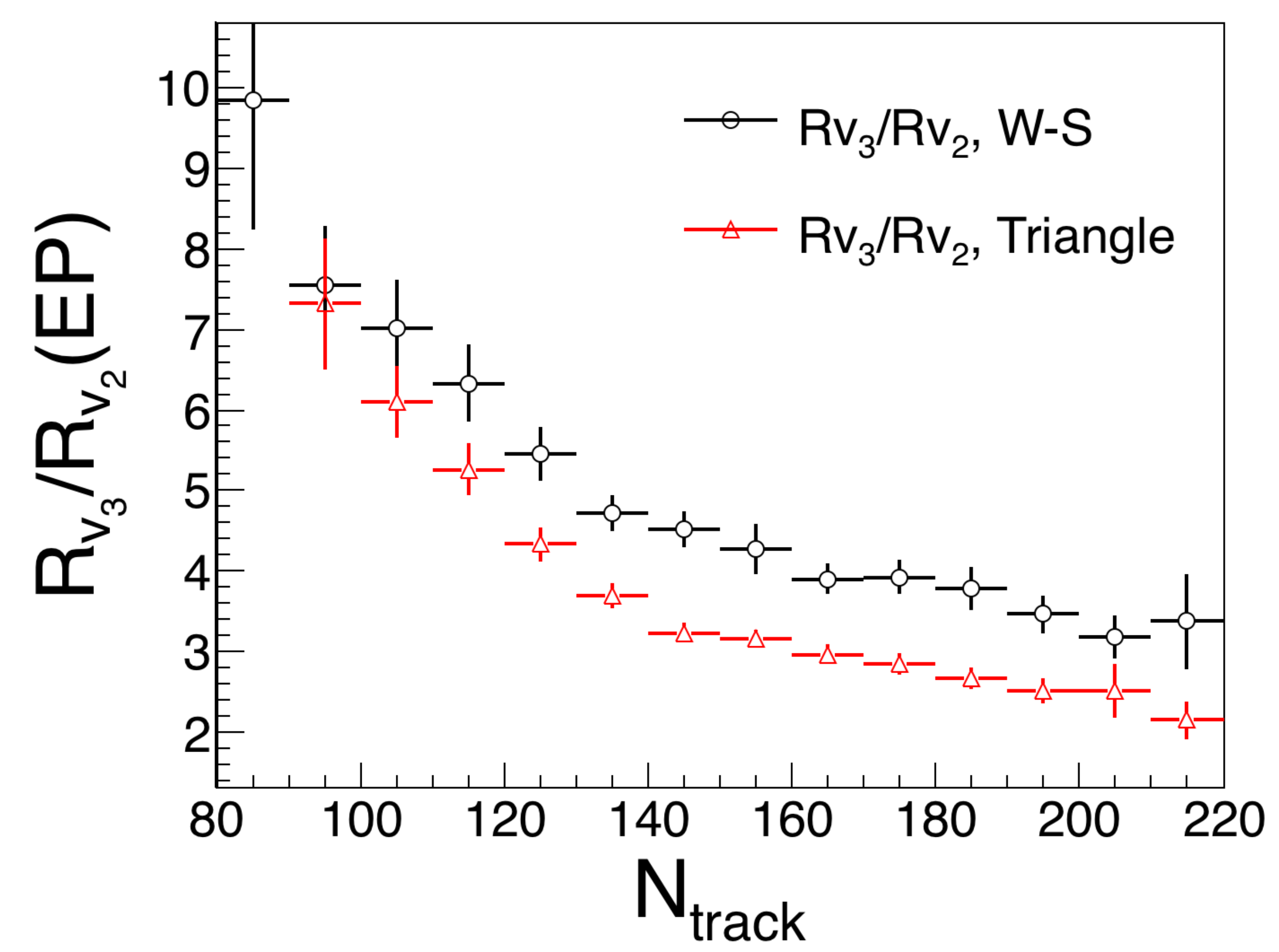}}
\caption{(Color online) Ratio of the relative flow fluctuation $R_{v_{3}}$/$R_{v_{2}}$ as a function of $N_{track}$ for $^{12}$C+$^{197}$Au collisions with $\alpha$-clustered $^{12}$C in different structures. Flow fluctuations are calculated based on the event-plane method.}
\label{f6}
\end{figure}

We further study the ratio of flow fluctuations. Fig.~\ref{f6} displays the ratio of the relative fluctuation of $v_{3}$ and $v_{2}$. Significant difference in magnitude can be seen for the $N_{track}$ dependence of $R_{v_{3}}$/$R_{v_{2}}$ for W-S and Triangle $\alpha$-clustering configurations. It is found that both Triangle and W-S configurations show monotonic $N_{track}$ dependence. The ratio is more sensitive for distinguishing Triangle and W-S configurations of $^{12}$C at large $N_{track}$.

\begin{figure*}[htbp]
\centering
\resizebox{15.0cm}{12.0cm}{\includegraphics{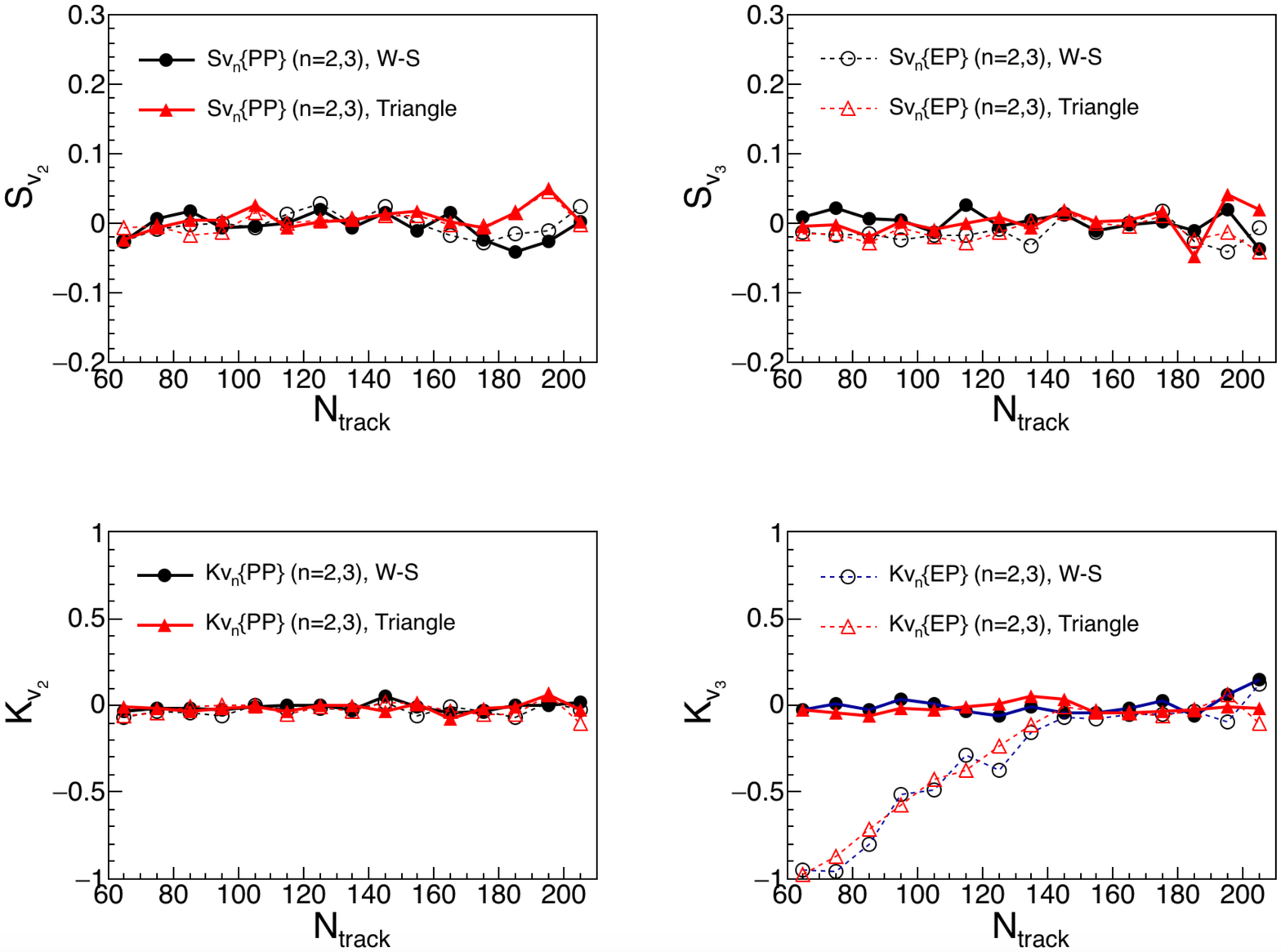}}
\caption{(Color online) Skewness and kurtosis of $v_{2}$ and $v_{3}$ fluctuation as a function of $N_{track}$ for $\alpha$-clustered $^{12}$C+$^{197}$Au collisions at 200 GeV. Comparisons are made between results from event plane method and participant plane method. Upper panels: Skewness of $v_{2}$ and $v_{3}$ fluctuation. Lower panels: Kurtosis of $v_{2}$ and $v_{3}$ fluctuation.}
\label{f7}
\end{figure*}

Assuming linear flow response, the approximation relation shown by Eq.(~\ref{q9}) also applies to the kurtosis and skewness of flow fluctuations. We study the skewness ($S_{v_{n}}$) and kurtosis ($K_{v_{n}}$) of $v_{n}$ ($n$=2,3) fluctuation in a similar way as corresponding $\varepsilon_{n}$ fluctuation as approaches for investigating the initial $\alpha$-clustering effect. Results from both the participant plane method and the event plane method are shown in comparison in Fig.~\ref{f7}. Different from the eccentricity fluctuation, one can find that both skewness and kurtosis of the flow harmonic are insensitive for distinguishing the unclustered W-S configuration and the triangular $\alpha$-clustering configuration. It could be from the source evolution that in the final-state of AMPT the $v_{n}$ ($n$=2,3) distributions are nearly Gaussian, thus the skewness and kurtosis of elliptic and triangular flow fluctuations especially at large $N_{track}$ are almost consistent with zero within statistical uncertainties.  

\subsection{Correlation of the initial eccentricity with final flow harmonic}

Impressive progresses have been made in studying the sensitivity of flow response to the initial geometry in relativistic heavy-ion collisions~\cite{Petersen2012,Niemi2013}. We understand that elliptic flow $v_2$ and triangular flow $v_3$ are driven mainly by the linear response to the initially produced ellipticity and triangularity of the source geometry. For asymmetric colliding systems involving $\alpha$-clustered nucleus, the conversion of the initial geometry to the final flow could become more complicated. Quantitative study of the initial-final correlation in event-by-event basis in model simulation is important for understanding the sensitivity of flow and flow fluctuation in probing clustering configurations. 

In particular, we study the correlations of the initial eccentricities with final flow harmonics for $^{12}$C+$^{197}$Au collisions with $\alpha$-clustered $^{12}$C configurations. Pearson coefficient is used to quantify the strength of the correlation. We define the coefficient in the following form which takes into account both the magnitude and the angle:

\begin{equation}
C_{v_{n},\varepsilon_{n}}=\frac{\langle v_{n}\varepsilon_{n} cos(n[\psi-\psi_{PP}]) \rangle}{\sqrt{\langle |\varepsilon_{n}|^{2} \rangle \langle |v_{n}|^{2} \rangle}},
\label{q10}
\end{equation}
where $\psi_{PP}$ is the participant plane angle, $\psi$ is the phase of the flow coefficient $v_{n}$ (e.g. $\psi$ is the event-plane angle if $v_{n}$ is calculated based on event plane method). As $C_{v_{n},\varepsilon_{n}}$ approaches one, good linear correlation can be expected, whereas $C_{v_{n},\varepsilon_{n}}$ approaches zero other contributions beyond linear response could contribute. We remark here the Pearson coefficent can be used to quantify correlations even non-linear effect contributes.

\begin{figure*}[htbp]
\centering
\resizebox{15.0cm}{!}{\includegraphics{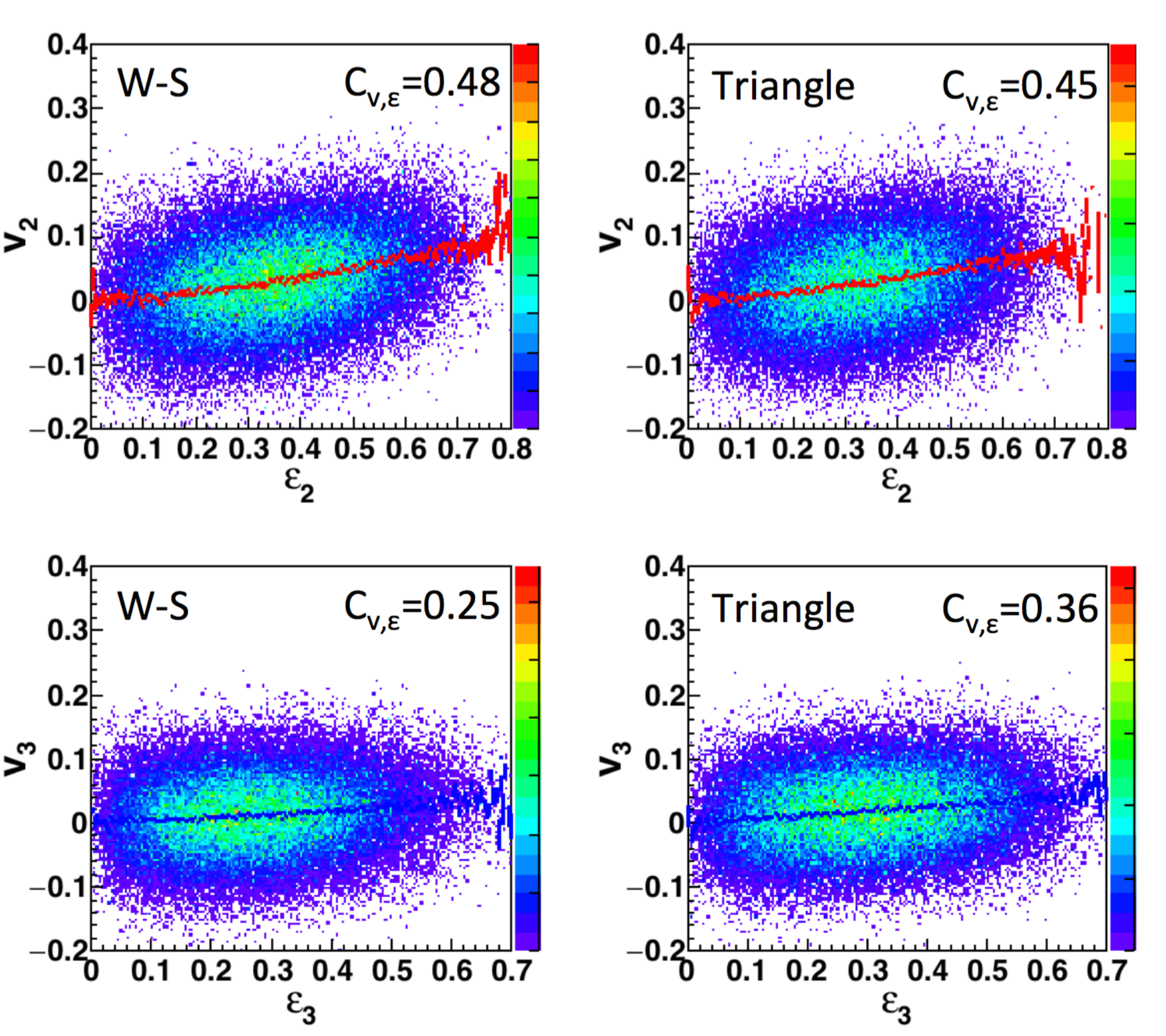}}
\caption{(Color online) 2-D plots illustrating the event-by-event correlation of $v_{n}$ and $\varepsilon_{n}$ for $^{12}$C+$^{197}$Au collisions with W-S and Triangle $\alpha$-clustering configuration of $^{12}$C. The values of the correlation coefficients $C_{v_{n},\varepsilon_{n}}$ are shown in the plots. Upper panels: $v_{2}$ vs $\varepsilon_{2}$ for 80$<$$N_{track}$$<$100. Lower panels: $v_{3}$ vs $\varepsilon_{3}$ for 80$<$$N_{track}$$<$100. Red and blue data points present the $\varepsilon_{n}$-bin averaged profile.}
\label{f8}
\end{figure*}

The two-dimensional plots in Fig.~\ref{f8} show the correlations between $v_{n}$ and $\varepsilon_{n}$ calculated using participant plane method. As can be seen in the figures, for collision events with 80$<$$N_{track}$$<$100, the $v_{2}$ and $v_{3}$ coefficients display a strong linear correlation to their corresponding initial-state coefficients for all the $\alpha$-clustering cases considered.

\begin{figure*}[htbp]
\centering
\resizebox{15.0cm}{!}{\includegraphics{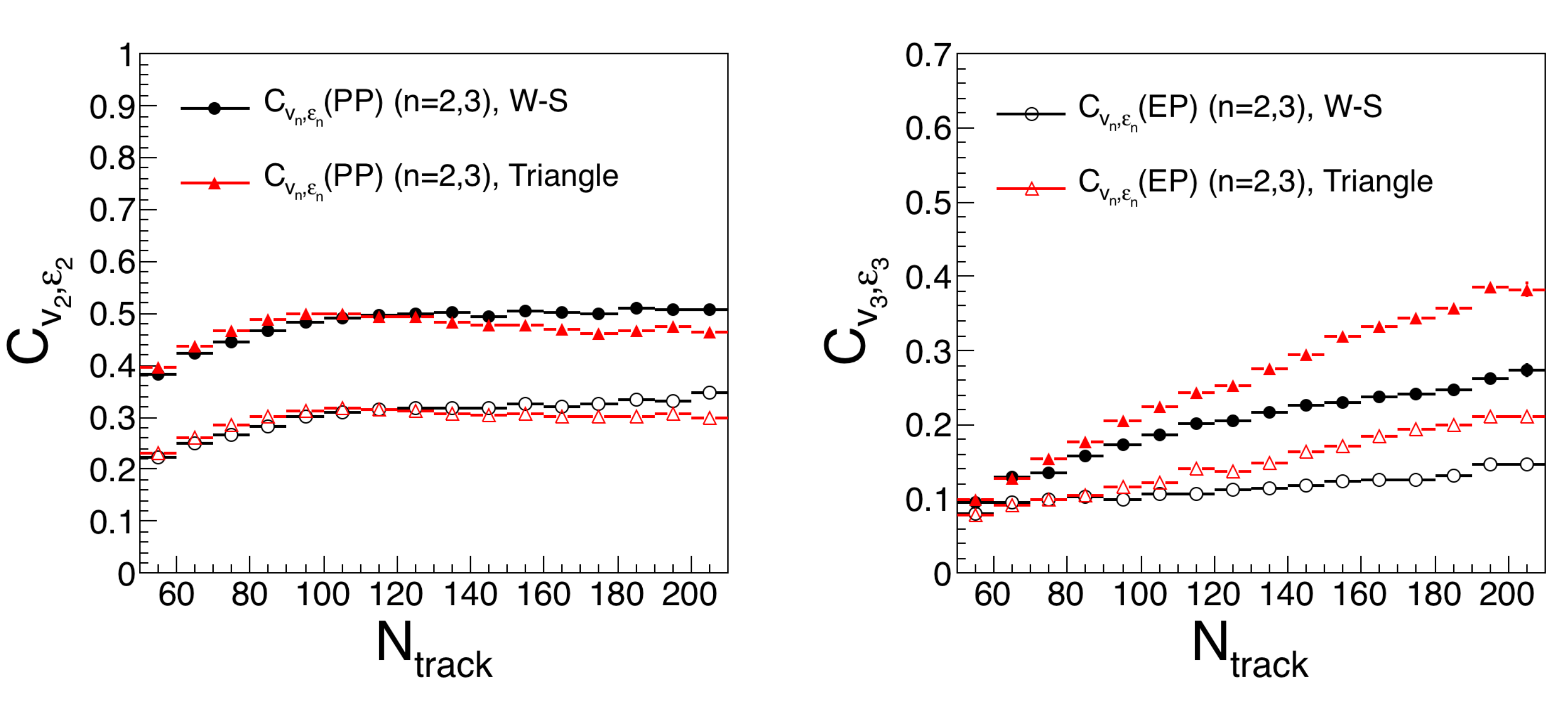}}
\caption{(Color online) Correlation coefficients $C_{v_{n},\varepsilon_{n}}$ (n=2,3) as a function of $N_{track}$ for $^{12}$C+$^{197}$Au collisions with different $^{12}$C $\alpha$-clustering configurations.}
\label{f9}
\end{figure*}

Fig.~\ref{f9} compares the model predictions of the $N_{track}$ dependence of the correlation function defined by Eq.(~\ref{q10}) for $^{12}$C+$^{197}$Au collisions with W-S and $\alpha$-clustered $^{12}$C configurations. $C_{v_{n},\varepsilon_{n}}$ ($n$=2,3) are generally seen to follow a smooth increasing trend as a function of $N_{track}$ indicating stronger linear $v_{n}$-$\varepsilon_{n}$ correlation at larger $N_{track}$. In comparison, W-S results are comparable with triangle configuration for the correlation between $v_{2}$ and $\varepsilon_{2}$ whereas triangle configuration presents stronger $v_{3}$-$\varepsilon_{3}$ correlation than W-S especially at large $N_{track}$.

\section{Summary}

In summary, we studied the $\alpha$-clustering effects on the initial eccentricity and final flow fluctuations in $\alpha$-clustered $^{12}$C+$^{197}$Au collisions at center-of-mass energy of 200 GeV using a multi-phase transport model.

Event-by-event fluctuations of the initial eccentricity $\varepsilon_{n}$ and final anisotropic flow $v_{n}$ are characterized by scaled standard variance, skewness and kurtosis. Differences in the multiplicity dependence of the flow fluctuation are observed for collision systems with W-S and Triangle $\alpha$-clustering configurations of $^{12}$C. The triangular flow fluctuation is shown to have particular sensitivity in distinguishing the triangle $\alpha$-clustering configuration which is consistent with the picture of the flow response to the initial geometry. The ratio of the triangular flow fluctuation over elliptic flow fluctuation decreases with increasing $N_{track}$ for both W-S and triangle configurations but show difference in magnitude. We also note that the differences in flow fluctuation between clustered and unclustered W-S structures are more significant in high-multiplicity collisions. Furthermore, correlations between flow harmonic and initial eccentricity are investigated. The correlation functions for $^{12}$C+$^{197}$Au collisions generally show an increasing trend as a function of $N_{track}$, indicating stronger linear correlations at larger $N_{track}$. In the picture of the flow response to the initial geometry, it is expected that harmonic flow or flow fluctuation can better probe the $\alpha$-clustering structure in higher multiplicity collision events.

Experimentally, as the ground state of $^{12}$C could exist as a configuration mixing state in reality, the actual significance of the fluctuation observable in distinguishing clustered configurations could be lower than we obtained in the ideal cases in our study. Nevertheless, because of the particular sensitivity of the flow fluctuation in reflecting the clustering structure, one could take flow fluctuation in addition to the flow measurement as an effective probe for the $\alpha$-clustering structure. 

As in high energy nuclear collisions, study of the event-by-event anisotropy fluctuation is crucial for understanding not only system initial conditions but also the afterburner evolution properties, future studies by looking into the flow fluctuation in other light-heavy collision systems (e.g. $^{16}$O+$^{197}$Au, $^{16}$O+$^{208}$Pb) with $\alpha$-clustering configurations will be promising for providing more important information about the nuclear clustering effect.

\section*{Acknowledgements}
This work is supported in part by the National Natural Science Foundation of China under contract Nos. 11890714, 11421505, 11905034, 11925502, 11935001, 11961141003, the Key Research Program of Frontier Sciences of the CAS under Grant No. QYZDJ-SSW-SLH002 and the Key Research Program of the CAS under Grant No. XDPB09.

\bibliographystyle{apsrev.bst}

\end{CJK*}
\end{document}